\documentclass[prd,twocolumn,showpacs,floatfix]{revtex4}
\usepackage{bm}
\usepackage{amssymb}
\usepackage{graphicx}
\usepackage{epstopdf}

\begin{document}
\title{Accurate time--domain gravitational waveforms for extreme-mass-ratio binaries}
\author{Lior M.~Burko}
\affiliation{Department of Physics, University of Alabama in Huntsville,
Huntsville, Alabama 35899, USA} 
\author{and Gaurav Khanna}
\affiliation{Physics Department,
University of Massachusetts at Dartmouth, N. Dartmouth, Massachusetts 02747, USA }
\date{December 20, 2006}
\begin{abstract}
The accuracy of time-domain solutions of the inhomogeneous Teukolsky equation is improved significantly. Comparing energy fluxes in gravitational waves with highly accurate frequency-domain results for circular equatorial orbits in Schwarzschild and Kerr, we find agreement to within 1\% or better, which may be even further improved. We apply our method to orbits for which frequency-domain calculations have a relative disadvantage, specifically high-eccentricity (elliptical and parabolic) ``zoom--whirl" orbits, and find the energy fluxes, waveforms, and characteristic strain in gravitational waves.  
\end{abstract}
\maketitle

\section{Introduction}

A stellar mass compact object (a black hole or a neutron star) moves in the spacetime of a supermassive black hole in an (accelerated) orbit, emitting low-frequency gravitational waves in the good sensitivity band of space borne detectors such as LISA. In this Paper we model a likely scenario in galactic nuclei, specifically the capture of compact stellar-size objects by a supermassive black hole, like those that exist in the center of many galaxies. Because of the high mass ratio, the compact object may be approximated by a point particle, and the orbital evolution modeled by perturbation theory. 

The emission of gravitational waves by a point-particle falling into a black hole has been modeled by several researchers using a 
frequency-domain (FD) decomposition of the master equation (see, e.g., \cite{FT, Hug} and references cited therein) \cite{sopuerta}. The FD approach allows for very accurate determination of the energy flux and the waveforms for wide classes of orbits that include orbits characterized by low values of the eccentricity. Generic orbits and especially high eccentricity orbits, although in principle amenable to a Fourier decomposition and a FD construction of the waveforms \cite{schmidt,Hug}, require the summation of many terms in the Fourier series, which limits the accuracy and increases the computation time \cite{FD-comment}. This weakness of the FD approach is not shared with the time-domain (TD) approach~\cite{Ramon,Gaurav,Karl} that we present here. Indeed, wave generation is typically one of the strengths of TD approaches. We expect a TD approach to be more efficient and yield more accurate results (for waveforms) especially in the context of highly elliptic and parabolic orbits compared with FD calculations. TD calculations can also serve as an independent test of FD results. 

Calculations in the TD have only recently been introduced for the cases of equatorial circular particle orbits~\cite{Ramon} and also for equatorial  elliptic orbits and inclined circular orbits~\cite{Gaurav} around a rapidly rotating black hole. These calculations, however, suffer from low accuracy, as was manifest in deviations by as much as $10$--$20\%$ in the determination of the energy flux when compared with the highly accurate FD results. Indeed, it is this crudity of TD calculations that was the main reason why so few TD calculations have been done. Here, we report on a significant improvement in the accuracy of TD calculations, that brings them to within $1\%$ agreement with the FD counterparts or better. Then, we apply our approach to a special class of orbits, namely ``zoom--whirl'' orbits of both elliptic  \cite{hug3} and parabolic type. As was recently pointed out in \cite{babak}, adiabatic waveforms that ignore the gravitational wave backreaction on the orbit may yield so-called ``kludge" waveforms, that are surprisingly accurate for much of the parameter space of extreme-mass-ratio binary motion. We adopt here a similar approach, and neglect the effects of radiation reaction, hoping to address them elsewhere. The purpose of this Paper is to serve as a {\it proof-of-concept} that TD calculations of gravitational waveforms can be accurate---specifically where FD calculations have weaknesses---and hence be of practical importance for waveform generation. This Paper is not intended to serve as a reference for tables of results, or to exhaustively cover the relevant parameter space. The latter---in addition to additional improvements to the code and model---will be presented elsewhere. 

This paper is organized as follows. In Section \ref{s2} we briefly review the TD approach. In  Section \ref{s3} we demonstrate the greatly improved accuracy of our approach by reproducing results obtained with FD calculations to within $1\%$. Next, in Section \ref{s4}, we present new results for accurate TD waveforms for equatorial elliptic and parabolic, zoom--whirl orbits. Finally, in Section \ref{s5} we discuss our results. 

\section{Time-domain calculations}\label{s2}

The Teukolsky equation, that governs the perturbations of Kerr geometry in Boyer-Lindquist coordinates with a matter-source term, is given by \cite{Teuk}
\begin{eqnarray} \label{te}
&&{}\left [ \frac{(r^2+a^2)^2}{\Delta} - a^2 \sin^2 \theta \right ]
\frac{\partial^2 \psi}{\partial t^2} + \frac{4 M a r}{\Delta}
\frac{\partial^2 \psi}{\partial t
\partial \phi} \nonumber \\
&&+ \left [ \frac{a^2}{\Delta} - \frac{1}{\sin^2 \theta} \right ]
\frac{\partial^2 \psi}{\partial \phi^2}
- \Delta^{-s} \frac{\partial}{\partial r} \left ( \Delta^{s+1}\frac{\partial
\psi}{\partial r}\right ) \nonumber \\
&&
- \frac{1}{\sin \theta} \frac{\partial}{\partial \theta} \, \left ( \sin
\theta \frac{\partial \psi}{\partial \theta} \right )
-2 s \left [\frac{a (r-M)}{\Delta} + \frac{i
\, \cos \theta}{\sin^2 \theta} \right ] \frac{\partial
\psi}{\partial \varphi}  \nonumber \\ 
&& - 2 s \left [ \frac{M(a^2 - r^2)}{\Delta} -
r - i\, a\, \cos \theta \right ] \frac{\partial
\psi}{\partial t} \nonumber \\
&& + \left [ s^2 \cot^2 \theta - s \right ] \psi = 4 \pi (r^2+a^2\cos^2\theta) T
\end{eqnarray}
where $\Delta=r^2-2M r+a^2$ and $T = 2 \rho^{-4}\; T_4$,
where
\begin{eqnarray}\label{T4}
T_4 &=& (\Delta + 3 \gamma - \gamma^* + 4 \mu + \mu^*)[(\delta^*
- 2 \tau^* + 2 \alpha)T_{nm^*} \nonumber \\ &-& (\Delta + 2
\gamma - 2 \gamma^* + \mu^*)T_{m^*m^*}] + (\delta^* - \tau^* +
\beta^* + 3 \alpha \nonumber \\
&+& 4 \pi) [(\Delta + 2
\gamma + 2 \mu^*)T_{nm^*} \nonumber \\ &-& (\delta^* - \tau^* + 2 \beta^* + 2
\alpha) T_{nn}].
\end{eqnarray}
All the symbols used above are defined in \cite{Teuk}. By choosing their values in Boyer-Lindquist coordinates and expressing Eq.~(\ref{te}) explicitly, we obtain an explicit but quite complicated form for the source term. We treat the delta functions that appear in (\ref{T4}) by using a ``regularization'' in which we substitute the delta function with a very narrow Gaussian function (represented only by a few grid points),
\begin{equation}\label{gauss}
\delta(x - x(t))\,\approx\,\frac{1}{\sqrt{2
\pi}\,\sigma}\,\exp\left(\frac{-(x-x(t))^2}{2
\sigma^2}\right)
\end{equation}
for small $\sigma$. We handle the deltas in the radial and angular direction through the Gaussian approximation, whereas the $\delta(\phi - \phi(t))$ function is handled analytically. The methodology for numerically evolving the homogeneous part of (\ref{te}) has been presented in detail \cite{Laguna}. We take a similar approach, except that we include the source term which is essential for modeling extreme-mass-ratio binaries. More details can be found in \cite{Ramon}.

Note that we take the initial data for the fields to be zero. Since we are considering the Teukolsky equation with a source, this choice corresponds to the particle ``appearing suddenly,'' which generates an artificial burst of spurious radiation, that decays fast. We handle this difficulty by evolving to late times and computing fluxes only after the system has ``settled down.'' We used the highly accurate public domain integrator  ``Geod''  \cite{Geod}---which we call from within our Teukolsky code---to update the particle's position at every time step.  

Finally, to obtain the power (``total energy flux") due to gravitational wave emission from the binary, we find the  Teukolsky function $\psi$ at a set radial location in the grid far from the horizon, and compute \cite{ djdt1}
\begin{eqnarray}
\frac{{\rm d} E}{{\rm d} t}&=&\lim_{r\to\infty}\left[\frac{1}{4
\pi r^6} \int_{\Omega} \,{\rm d}\Omega \left| \int_{-\infty}^{t}
\,{\rm
d}\tilde{t}\,\psi(\tilde{t},r,\theta,\varphi)\right|^{2}\,\right]\, .
\label{jfl}
\end{eqnarray}
The characteristic strain in gravitational waves is then estimated by
\begin{equation}
h_c=\frac{\sqrt{2}(1+z)}{\pi\;D_{\rm L}(z)}\;\sqrt{\left| \frac{\,dE_e}{\,df_e}\right|}\, ,
\end{equation}
where $z$ is the cosmological redshift, $D_{\rm L}$ is the luminosity distance, and $|\,dE_e/\,df_e|$ is the total energy per unit frequency carried away by the gravitational waves (in the ``emitter's'' reference frame, hence the subscripts ``e").

\section{Testing the code's accuracy}\label{s3}

The numerical implementation of this work is in the form of a set of 2+1 dimensional linear PDE's that uses the Lax--Wendroff evolution scheme. This approach, including convergence and stability, is described in \cite{Ramon}.

We test the accuracy of our code by performing standard convergence tests, and we find the code to converge with second order. We also tested the code by comparing our results for the energy flux in gravitational waves with those obtained with FD calculations. Specifically, we compare circular and equatorial orbits for Schwarzschild and Kerr black holes, where the FD fluxes are known to high accuracy.

As discussed above, earlier TD calculations suffered from crude accuracy, with energy fluxes typically deviating from their FD counterparts by 10--20\%. Analysis of the reasons for that crudity allowed us to attribute it primarily to the relatively coarse grids used in \cite{Ramon,Gaurav}, and to the relatively small size of the computational domain, which required the evaluation of the field only a small number of wavelengths from the source, in addition to unoptimized modeling of the point particle. Moreover, there exists a typographical sign error in the expression for $C_{{\bar m},n}$ in Ref.~\cite{Dan} which was used in \cite{Ramon,Gaurav} for computing the source-term. This wrong sign introduces a significant error in the strong-field region of the black hole space-time, but has little impact when orbits with large radii are considered. Finally, it should also be noted that the extra normalization factor as used before (see Eq.~20 in~\cite{Ramon}) is actually inappropriate in the context of our computation. In our current code we have fixed the sign error as mentioned above and removed this improper normalization. Also, owing to certain optimization and performance improvements to the code, we can now have much larger grid sizes---and therefore extract the waveforms and compute the fluxes at a distance much farther than in \cite{Ramon,Gaurav}---and also have much higher resolution.

\begin{figure}
\input epsf
\includegraphics[width=8.0cm]{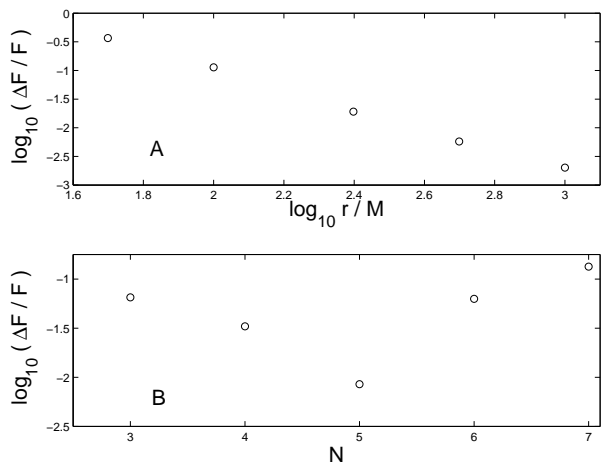}
\caption{The relative error in the energy flux in gravitational waves for a particle of mass $\mu/M=0.01$ in a circular equatorial orbit of radius $5r_{\rm isco}$ about a Kerr black hole with spin parameter $a/M=0.9$. Upper panel (A):  As a function of the distance at which wave extraction is done. The errors are calculated with a value corresponding to wave extraction at infinity, that we obtain using Richardson's extrapolations. Here, $N=5$. Lower panel (B): As a function of the number of points used to sample the Gaussian $N$. The errors are calculated with the FD value. Wave extraction is done at $500M$. Data for both panels are calculated at our chosen grid density.}
\label{ext-distance}
\end{figure}

%\begin{figure}
%\includegraphics[width=6.0cm]{m2ellip.eps}
%\caption{Waveform from the elliptic orbit for the dominant mode $m=2$. We display $Re (\psi)$ (solid curve) and $Im (\psi)$ (dotted curve) as functions of time. Note the ``initial burst'' which should be ignored. The smooth, slow oscillations from the ``whirl'' part of the orbit are clearly visible. The waveform drops off very rapidly in the ``zoom'' part of the orbit. }
%\end{figure}

In Fig.~\ref{ext-distance}A we show the relative error in the determination of the energy flux in gravitational waves as a function of the distance at which wave extraction is done, for a particle of mass $\mu/M=0.01$ in a circular equatorial orbit of radius $5r_{\rm isco}$ about a Kerr black hole with spin parameter $a/M=0.9$. In practice, we extract the waveforms and compute the fluxes at or beyond $500M$. Here, and similarly throughout this paper, the ``detector" is positioned on the equatorial plane of the black hole, although there are no restrictions in positioning it at any inclination angle with respect to the equatorial plane. 
Our typical grid resolutions are $0.025M$ (radial) and $0.05$ (angular). Modeling the particle by a Gaussian distribution requires optimization of the number of grid points $N$ that are used to sample the Gaussian: For a given grid resolution, too many grid points cause the physical width of the Gaussian to spread too much resulting in large deviations from the point-particle approximation. On the other hand, too few grid points cause the Gaussian to be ``under-sampled,'' introducing unwanted errors into the computation. Through extensive numerical experimentation, we have found the optimal number of grid points across the gaussian for our chosen grid resolution to be about five. Figure 
\ref{ext-distance}B shows the relative errors in the determination of the flux extracted at $500M$ as a function of $N$, for the same physical system. For the modes $m=2,3$ that we report here, we find that $N=5$ is optimal. This value for $N$ may change for other values for $m$. 
We also experimented in modeling the particle with non-Gaussian profiles, e.g., high-order polynomials with compact support, and a ``discrete-delta'' approach werein one starts out by defining a step function on a numerical grid and takes finite-differenced derivatives
to obtain an approximation to the delta function and its derivatives. Done optimally, the results one obtains from these various implementations are consistent. Further details about the ``discrete-delta" approach will be discussed elsewhere. We consider this as one of the strengths of our approach: our results are quite robust.

\begin{table}
 \caption{Comparison of gravitational wave total energy fluxes detected at infinity using the FD solution, with the results measured numerically  at $r = 500 M$ while evolving the Teukolsky equation in the TD. The particle of mass $\mu/M=0.01$ is in a circular equatorial orbit of radius $r=r_{\rm isco}$ about a Schwarzschild black hole.}
  \centering
     \begin{tabular}{|c|c|c|} \hline
   {\bf m mode} & {\bf Time domain} & {\bf Frequency domain}\cr
    & {\bf energy flux} & {\bf energy flux }
   \cr \hline \hline
    $2$ &      $7.31\times10^{-8}$ &  $7.37\times10^{-8}$ \cr \hline
    $3$ &      $1.45\times10^{-8}$ &  $1.46\times10^{-8}$ \cr \hline
    $4$ &      $3.57\times10^{-9}$ &  $3.60\times10^{-9}$ \cr \hline   
   \end{tabular}
\label{comp1}
\end{table}

\begin{table}
 \caption{Same as Table \ref{comp1}, for a particle of mass $\mu/M=0.01$ in a circular equatorial orbit of radius $5.0\, r_{\rm isco}$ about a Kerr black hole with spin parameter $a/M = 0.9$.}
  \centering
     \begin{tabular}{|c|c|c|} \hline
   {\bf m mode} & {\bf Time domain} & {\bf Frequency domain}\cr
    & {\bf energy flux} & {\bf energy flux }
   \cr \hline \hline
    $2$ &      $2.19\times10^{-9}$ &  $2.21\times10^{-9}$ \cr \hline
    $3$ &      $2.18\times10^{-10}$ &  $2.20\times10^{-10}$ \cr \hline
    $4$ &      $2.68\times10^{-11}$ &  $2.71\times10^{-11}$ \cr \hline   
   \end{tabular}
\label{comp2}
\end{table}

%\begin{figure}
%\includegraphics[width=6.0cm]{m3ellip.eps}
%\caption{Same as Fig.~3, for the mode $m=3$.}
%\end{figure}

Implementing these refinements in the code, we considerably improve the accuracy of our TD results compared with the previous standard. Sample comparisons with FD are described in tables \ref{comp1} and \ref{comp2}. Our results agree with FD results to about 1\% or better. We anticipate that this agreement can be further improved, by an additional increase in resolution (possibly using a multi-grid approach of extrapolation to infinite resolution), extracting the field farther out (again, with a possible extrapolation to infinity), and by more experimentation with various models for the point particle.

\begin{figure}
 \includegraphics[width=8.0cm]{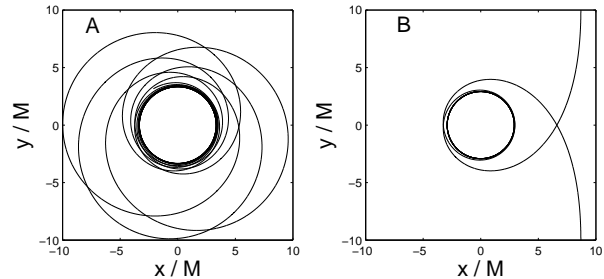}
%\begin{figure}
%\epsfysize=80mm \epsfbox{orbits.eps} 
\caption{Equatorial zoom--whirl orbits used in the text. Left panel (A): An elliptic orbit about a Kerr black hole with spin $a/M=0.5$, with $p=5.0\,M$ and  $e=0.5$. This orbit includes a ``whirl'' part that has about four full quasi-circular cycles, before it enters the ``zoom'' phase. Right panel (B): A parabolic orbit about a Kerr black hole with spin $a/M=0.5$, with $p=5.828427\,M$ and $e=1$. This orbit includes a ``whirl'' part that has about eight full quasi-circular cycles, before it enters the ``zoom out'' phase.}
\label{fig1}
\end{figure}

%\begin{figure}
 %\includegraphics[width=6.0cm]{ellip_orbit.eps}
%%\begin{figure}
%%\epsfysize=80mm \epsfbox{elliptic.eps} 
%\caption{An elliptic, zoom--whirl orbit about a Kerr black hole with spin $a/M=0.5$, with $p=5.0\,M$ and  $e=0.5$. This orbit includes a ``whirl'' part that has about four full quasi-circular cycles, before it enters the ``zoom'' phase. }
%\label{fig1}
%\end{figure}

%\begin{figure}
%\input epsf
%\includegraphics[width=6.0cm]{parab_orbit.eps}
%\caption{A parabolic zoom--whirl orbit about a Kerr black hole with spin $a/M=0.5$, with $p=5.828427\,M$ and $e=1$. This orbit includes a ``whirl'' part that has about eight full quasi-circular cycles, before it enters the ``zoom out'' phase. }
%\label{fig2}
%\end{figure}

\begin{figure}
\includegraphics[width=8.5cm]{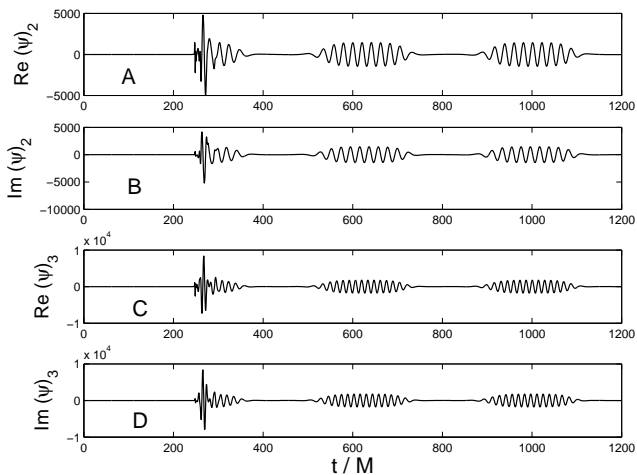}
\caption{Waveform from the elliptic orbit. We display ${\rm Re} (\psi)$ and ${\rm Im} (\psi)$ as functions of time. Note the ``initial burst'' which should be ignored. The smooth, slow oscillations from the ``whirl'' part of the orbit are clearly visible. The waveform drops off very rapidly in the ``zoom'' part of the orbit. Panel A: Real part of the dominant mode ($m=2$). Panel B: Imaginary part of the same. Panel C: Real part of the mode $m=3$. Panel D: Imaginary part of the same.}
\end{figure}

\section{Equatorial zoom--whirl orbits}\label{s4}

``Zoom--whirl'' orbits have been studied in detail~\cite{hug3}. The basic idea is that for certain elliptic orbits (close to the separatrix between stable and unstable orbits) the amount of time spent close to the periastron is very large, so that the particle traces a quasi-circular path close to the periastron before traveling back to the apastron. For high-eccentricity orbits, the particle appears to ``zoom in'' towards its periastron location, spend a number of circular revolutions (``whirls''), and then ``zoom out'' towards the apastron. These orbits are very interesting, especially for the case of rapidly spinning black holes, because of the large number of ``whirls'' they perform close to the hole. Similarly, there are parabolic orbits that exhibit similar ``whirl'' and ``zoom out'' characteristics. 

\begin{figure}
\includegraphics[width=8.0cm]{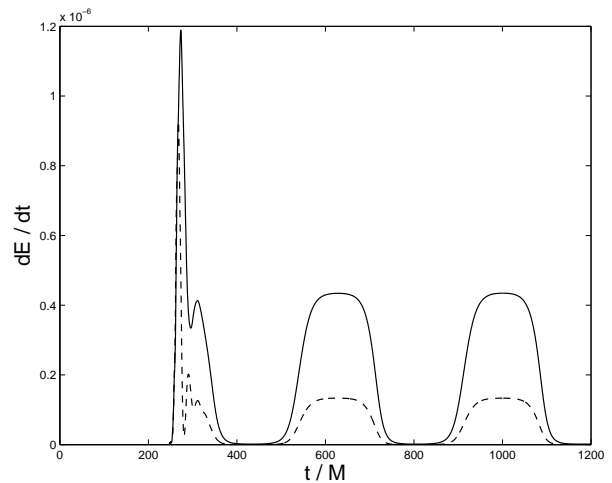}
\caption{Total energy flux from the elliptic orbit for the dominant mode $m=2$ (solid curve) and the $m=3$ mode (dashed curve). Again, the ``initial burst'' should be ignored. The almost steady flux from the ``whirl'' part of the orbit is clearly visible.}
\end{figure}

\begin{figure}
\includegraphics[width=8.0cm]{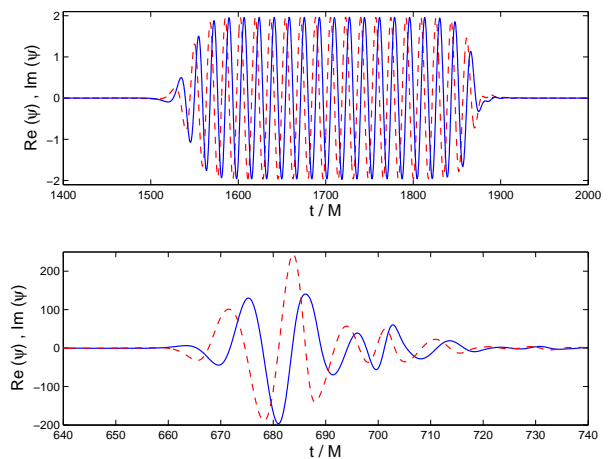}
\caption{Waveform from the parabolic orbit for the $m=2$ mode (upper panel). We display ${\rm Re} (\psi)$ (solid curve) and ${\rm Im} (\psi)$ (dashed) as functions of time. 
The spurious ``initial burst'' is plotted in the lower panel. The smooth, slow oscillations from the ``whirl'' part of the orbit are clearly visible. The waveform drops off very rapidly in the ``zoom out'' part of the orbit. }
\end{figure}

First, we plot sample equatorial zoom--whirl orbits about a Kerr black hole with spin $a/M=0.5$ in 
Fig.~\ref{fig1}A (elliptic orbit with semi-latus rectum $p=5.0\,M$ and eccentricity $e=0.5$) and in Fig.~\ref{fig1}B (parabolic orbit with $p=5.828427\,M$ and $e=1$).

We next show the waveforms and energy fluxes from a particle moving in the orbits depicted in Figs.~2A (in Figs.~3--4, henceforth the ``elliptic orbit") and 2B (in Figs.~5--7, hereafter the ``parabolic orbit"). The waveforms presented are the real and imaginary parts of the $m=2, 3$ modes of the Teukolsky function. All waveforms are extracted at $500M$ (i.e., these are the waveforms for an observer at that distance.) As mentioned above, the initial burst of radiation is spurious, as it corresponds to the particle ``appearing out of nowhere.'' The interesting features of these waveforms are as follows: The emitted waves become very prominent as the particle approaches the periastron, and fall off rapidly when the particle moves away from the hole. Note that the part of the waveform from the ``whirl'' seems to be similar to one coming from an exact circular orbit of the same  angular velocity. In fact, we checked this quantitatively, and indeed, the waveform of the ``whirl'' part exactly matches with the waveform that arises from a corresponding circular orbit. This happens both for the elliptic and parabolic case.

\begin{figure}
\includegraphics[width=8.0cm]{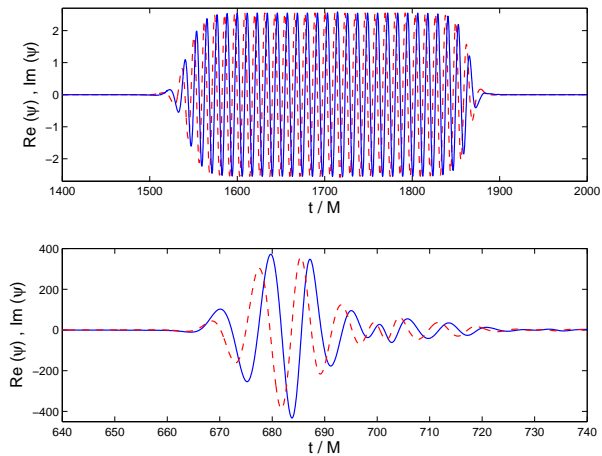}
\caption{Same as Fig.~5, for the mode $m=3$. }
\end{figure}

\begin{figure}
\includegraphics[width=8.0cm]{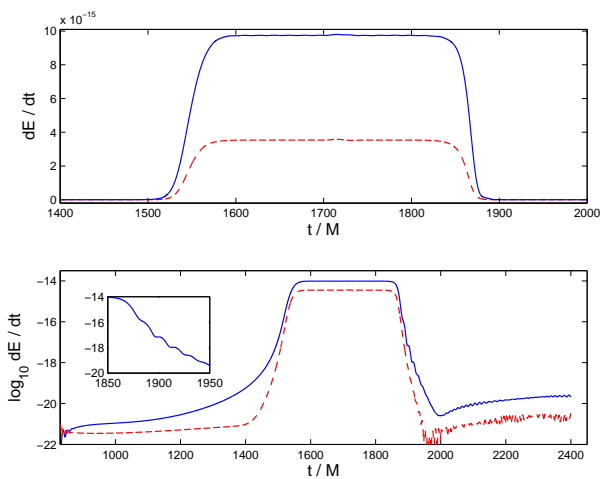}
\caption{Total energy flux from the parabolic orbit for the dominant mode $m=2$ (solid curve) and the $m=3$ mode (dashed curve). The initial burst was excised. The almost steady flux from the ``whirl'' part of the orbit is clearly visible.}
\end{figure}

\begin{figure}
\includegraphics[width=8.0cm]{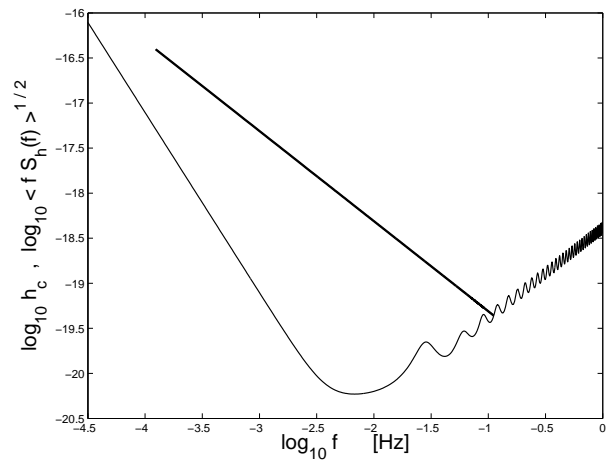}
\caption{Characteristic strain (bold curve) in gravitational waves $h_c$ for a $1M_{\odot}$ black hole moving on the parabolic orbit around a $10^6M_{\odot}$ black hole with $a/M=0.5$, at a distance of 1Gpc, shown on a standard LISA noise curve with signal-to-noise ratio of 1 \cite{lisa}.}
\end{figure}

It is clear from the figures that the physically significant parts of zoom--whirl orbits, as expected, are the ``whirls'' because the emitted radiation from that part is much stronger. This is interesting, because this implies that with some care, one may be able to estimate radiative fluxes etc.~for such orbits using known circular orbit results. 
Notice that although the parabolic orbit is symmetric (about the $x$-axis, i.e., the ``zoom in" phase of he orbit is symmetrical to the ``zoom out" phase), the waveforms and the energy flux are not symmetrical. Specifically, notice the oscillations in (the lower panel of) Fig.~7, that are visible only during the ``zoom out" phase. The reason for this asymmetry is that during the ``zoom out" phase of the orbit, the waves include also the waves emitted during the ``zoom in" and the ``whirl" phases, that backscatter off the curvature of spacetime. Indeed, the (real part of the) angular frequency of the oscillations in Fig.~7 is found numerically to equal $\sim 0.4\,M^{-1}$, which is very close to twice the orbital angular frequency of $0.183\,M^{-1}$. (Note that the decay of the ``initial burst" is characterized by oscillations with the quasinormal mode frequency.) Finally, we show in Fig.~8 the characteristic strain in gravitational waves $h_c$ for the parabolic orbit, for the dominant mode $m=2$. 

Notice that while for circular orbits (both Schwarzschild and Kerr) the energy flux in gravitational waves in the dominant mode $m=2$ is greater than that in the next mode ($m=3$) by an order of magnitude, for the zoom--whirl orbits we have studied it is greater by only a factor of $\sim2$. This difference may have implications on the number of modes that needs to be computed.

The new results we are presenting here for zoom--whirl orbits can be expected to be accurate within a 1\% or better. However, as mentioned before, we expect to be able to reduce this error through higher resolution, better particle modeling,  and related ideas. 

\section{Conclusions}\label{s5}

A variety of particle orbits in the context of extreme-mass-ratio binaries can be effectively studied using Kerr black hole perturbation theory in the TD. We demonstrate the potential accuracy of the TD approach by getting very good agreement (within 1\% or better) in comparisons with FD results for radiated energy fluxes for a particle in a circular orbit.  This agreement may be further improved. 

We also obtained gravitational ``waveforms'' arising from particles in a special class of orbits, namely zoom--whirl orbits of both elliptic and parabolic type. This is the first study of Kerr parabolic orbits. We observed that the ``whirl'' part of these orbits dominates the emitted radiation and is very similar to the pattern of radiation arising from a circular orbit of the appropriate parameters. We also present the characteristic strain in gravitational waves for a parabolic zoom--whirl orbit. 
We ignored the effects of radiation reaction in this work, with the hope of addressing those in detail elsewhere.

The TD Teukolsky approach to extreme-mass-ratio inspiral is maturing, and no longer suffers from the poor accuracy problems that used to plague it.  One may expect interesting and useful results from it where the FD approach has disadvantages, in addition to being a useful independent check on FD results. 

\acknowledgments

GK would like to thank P.~Sundararajan for many useful discussions. GK acknowledges
the research support provided by the University of Massachusetts, Dartmouth and the Glaser Trust of New York. LMB was supported in part by NASA EPSCoR grant No.~NCC5--580.

\end{document}